\documentclass{elsarticle}
\usepackage{latexsym}
\usepackage[latin2]{inputenc}
\usepackage[hidelinks]{hyperref}

\usepackage{lineno}

\makeatletter
\def\ps@pprintTitle{%
	\let\@oddhead\@empty
	\let\@evenhead\@empty
	\def\@oddfoot{\centerline{\thepage}}%
	\let\@evenfoot\@oddfoot}
\makeatother

\title{The "Giant Virus Finder" Discovers an Abundance of Giant Viruses in the Antarctic Dry Valleys}

\date{}

\author{Csaba Kerepesi,$^{\rm a}$, Vince Grolmusz\,$^{\rm a, b}$\footnote{to whom correspondence should be addressed: grolmusz@pitgroup.org}
\\
\small $^{\rm a}$ PIT Bioinformatics Group, Eötvös University,\\
 \small Pázmány Péter stny. 1/C, H-1117 Budapest, Hungary\\
\small $^{\rm b}$ Uratim Ltd.,  H-1118 Budapest, Hungary}

\begin{document}

	\begin{abstract}
		The first giant virus was identified in 2003 from a biofilm of an industrial water-cooling tower in England. Later, numerous new giant viruses were found in oceans and freshwater habitats, some of them having even 2,500 genes. We have demonstrated their very likely presence in four soil samples taken from the Kutch Desert (Gujarat, India). Here we describe a bioinformatics work-flow, called the `` Giant Virus Finder'' that is capable to discover the very likely presence of the genomes of giant viruses in metagenomic shotgun-sequenced datasets. The new workflow is applied to numerous hot and cold desert soil samples as well as some tundra- and forest soils. We show that most of these samples contain giant viruses, and especially many were found in the Antarctic dry valleys. The results imply that giant viruses could be frequent not only in aqueous habitats, but in a wide spectrum of soils on our planet.
		\bigskip
		
	\end{abstract}

\maketitle
\date{}

\section{Introduction} 

The mere existence of the giant viruses \cite{Raoult2004, Colson2011,  Yau2011,Boyer2009a,Garza1995,Fischer2010,Philippe2013} still posts challenges to the definition of life: some authors argue that they should be considered as the members of the ``fourth domain of life'' \cite{Claverie2006,Colson2011a,Colson2011b}, while some others are arguing that this is not the case \cite{Legendre2012,Williams2011}. Nevertheless, the complex interactions of the genes of the amoeba hosts of giant viruses with the viral-, bacterial- and eukaryotic genes may be accounted for the genetic variability of numerous organisms \cite{Fischer2011a,Desnues2012b,Jachiet2014}. 

Because the amoeba hosts of most of these viruses live in aqueous environments, almost all of these viruses were discovered in ponds, oceans, lakes or industrial water-cooling towers. 

By analyzing the metagenomes of the soil samples of the Kutch Desert (Gujarat, India) \cite{Pandit2014}, we have shown the presence of giant viruses in this periodically flooded, salty and hot environment \cite{Kerepesi2014b}.

In the present work we re-analyzed a dataset published with the article \cite{Fierer2012}, describing the soil microbiota of 16 samples of diverse geographic locations, including the North-American prairie, the Chihuahuan- and the Mojave deserts in New Mexico and California, the Antarctic dry valleys, the Alaskan tundra, and several forests in tropical and temperate regions. The focus of the work of \cite{Fierer2012} was the thorough metagenomic analysis of 16 environmental samples for bacteria and archaea, enlightening phylogenetic- and functional annotation of the nucleotide sequences found. No detailed analysis was performed for viruses and viral genes. 

Applying our new Giant Virus Finder workflow, we have found DNA segments of giant viruses in the samples, implying the very probable presence of giant viruses in these diverse soils. 

\subsection{The Giant Virus Finder}

Here we describe a general workflow that we have developed for the task of finding giant virus nucleotide sequences in metagenomic samples. The workflow is called the ``Giant Virus Finder'', and it is a collection of scripts with carefully set parameters for BLAST-based searches \cite{Altschul2005} of short-read metagenomic data sets. The workflow is available at the address \url{http://pitgroup.org/giant-virus-finder}.

The workflow is presented in detail in the ``Methods'' section and in summary on Figure 2. We emphasize here three important features: 

\begin{itemize}
	\item[(\i)] We have prepared a list of giant viruses that takes into account only the genome or (if there is no complete genome deposited) sequence size: viruses with 300 kbp or longer genomes or sequences are the members of the list. Clearly, all of the known giant viruses are in the list, but some large viruses, usually not listed as "giants", are also there; e.g., the Canarypox virus, or some large bacteriophages. We note that the user of the method can easily adjust this 300 kbp threshold to arbitrary other value.   
	\item[(\i\i)] Our method searches for the whole short read (and not only the best aligned subsequence of the short read), taken from the metagenomic dataset, in the NCBI Nucleotide Collection (nt). This is an important point: if a giant virus is present in the sample, then some short reads come entirely from its genome. 
	\item[(\i\i\i)] The word size in the BLAST searches \cite{Altschul2005} are set cautiously: Too long word size in BLAST searches would not find highly scored non-giant virus sequences in the specificity validation step. Short word sizes, however, increase the precision and also the computational time considerably.  We have used $w=7$ word size in blastn search \cite{Altschul2005} (instead of the default $w=28$ word size in Megablast or the $w=11$ word size in blastn.) In a 16-core server, the running time was a little over four days.
\end{itemize}

\section{Results and discussion}

We have examined the metagenomes collected and deposited with the article \cite{Fierer2012} for the presence of nucleotide sequences characteristic of giant viruses. 

The summary of our results is given on Figure 1. A detailed list of the best hits with extremely good E-values are given in Table 1. 

\begin{table}[]
	
	\centering
	\scriptsize
	\caption{Best hits, ordered by the E-value, found by applying the Giant Virus Finder for the 16 metagenomes of \cite{Fierer2012}. {\bf Column 1: Read identifier} identifies the read and the results related to the read, deposited at \url{http://pitgroup.org/giant-virus-finder}. {\bf Column 2: MTG} relevant digits that identify the metagenome (MTG) published in \cite{Fierer2012} and deposited at the MG-RAST site. {\bf Column 3: Location of the MTG}: Geographic name of the source sample, without country and continent denotation. {\bf Column 4: E value}: in Phase 2, the smallest (i.e., best) E-value of the hits found. {\bf Column 5: Identity} the number of identical nucleotides in the best-aligned hit. {\bf Column 6: Putative taxa:} Assigned taxon using the top 20\% rule of the MEGAN LCA algorithm \cite{Huson2011}. }
	\medskip
	\begin{tabular}{|l|l|l|l|l|l|}
		\hline
		{\bf Read identifier} & {\bf MTG} & {\bf Location of the MTG} & {\bf E-value} & {\bf Identity} & {\bf Putative taxa}                   \\ \hline
		6:88:18701:16918      & 803       & Lake Bonney Valley        & 4e-30         & 91/100         & O.Lake phycodnavirus 1          \\ \hline
		6:47:2094:15918       & 902       & Lake Fryxell Valley       & 8e-26         & 87/99          & Mimiviridie {[}family{]}              \\ \hline
		7:99:13938:20909      & 904       & Wright Valley             & 3e-25         & 86/98          & P.bursaria Chlor.virus   \\ \hline
		4:84:16596:9047       & 876       & Bonanza Creek LTER        & 1e-24         & 87/100         & Mimiviridie {[}family{]}              \\ \hline
		4:2:19051:10732       & 876       & Bonanza Creek LTER        & 1e-23         & 86/100         & Mimiviridie {[}family{]}              \\ \hline
		4:114:18824:12821     & 876       & Bonanza Creek LTER        & 1e-23         & 86/100         & Mimiviridie {[}family{]}              \\ \hline
		4:46:3341:11752       & 876       & Bonanza Creek LTER        & 1e-22         & 84/98          & Mimiviridie {[}family{]}              \\ \hline
		6:81:6130:14704       & 803       & Lake Bonney Valley        & 2e-20         & 83/99          & Mimiviridie {[}family{]}              \\ \hline
		6:114:9759:15200      & 902       & Lake Fryxell Valley       & 3e-19         & 71/80          & Pandoravirus dulc./sal.           \\ \hline
		4:22:15009:3518       & 876       & Bonanza Creek LTER        & 3e-19         & 80/95          & Enterobact.{[}fam.{]}phage \\ \hline
		4:104:7691:17992      & 901       & Lake Bonney Valley        & 3e-19         & 83/100         & Enterobact.{[}ord.{]}phage   \\ \hline
		6:73:2193:17269       & 902       & Lake Fryxell Valley       & 4e-18         & 82/100         & Mimiviridie {[}family{]}              \\ \hline
		6:62:15221:2441       & 803       & Lake Bonney Valley        & 1e-17         & 72/84          & Mimiviridie {[}family{]}              \\ \hline
		6:66:10892:20320      & 902       & Lake Fryxell Valley       & 4e-17         & 76/91          & Mimiviridie {[}family{]}              \\ \hline
		6:89:6245:20070       & 900       & Garwood Valley            & 1e-16         & 81/98          & Mimiviridie {[}family{]}              \\ \hline
		6:114:12016:8378      & 902       & Lake Fryxell Valley       & 5e-16         & 74/89          & Mimiviridie {[}family{]}              \\ \hline
		4:22:17523:8570       & 876       & Bonanza Creek LTER        & 5e-16         & 75/91          & Mimiviridie {[}family{]}              \\ \hline
		6:79:15305:6160       & 872       & Chihuahuan Desert         & 5e-16         & 80/99          & Mimiviridie {[}family{]}              \\ \hline
		6:39:10664:8341       & 900       & Garwood Valley            & 6e-15         & 72/86          & Mimiviridie {[}family{]}              \\ \hline
		7:52:4423:10207       & 904       & Wright Valley             & 6e-15         & 60/67          & Mimiviridie {[}family{]}              \\ \hline
		7:16:9740:9012        & 904       & Wright Valley             & 6e-15         & 73/89          & Mimiviridie {[}family{]}              \\ \hline
		4:7:2721:12270        & 873       & Chihuahuan Desert         & 2e-15         & 66/75          & P.bursaria Chlor.virus   \\ \hline
		5:83:4473:7350        & 874       & Toolik Lake LTER          & 2e-15         & 65/75          & Mimiviridie {[}family{]}              \\ \hline
		5:42:4010:17638       & 899       & Duke Forest               & 2e-15         & 81/99          & C.roenbergensis virus         \\ \hline
		7:31:3572:1747        & 904       & Wright Valley             & 2e-15         & 74/90          & Moumovirus                            \\ \hline
	\end{tabular}
	
\end{table}

\begin{figure}[t]
	\centering
	\includegraphics[width=4.4in]{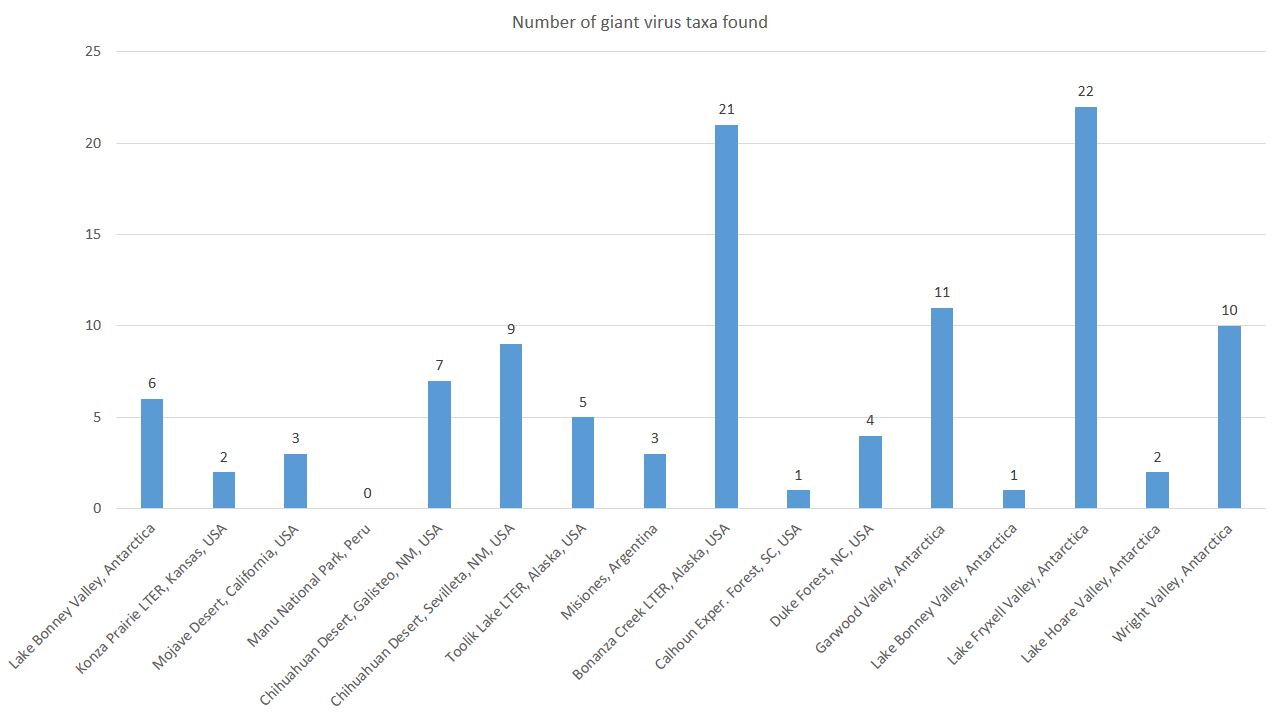}
	\caption{Summary of the results of the application of the Giant Virus Finder for the 16 metagenomes of \cite{Fierer2012}: Each metagenome is denoted on axis x by its geographic location, and the bars visualize the number of the giant virus reads found in the dataset. The MG-RAST accession number-labeled version of this figure is given as Figure S1 in the on-line supporting material. Detailed results can be found at \url{http://pitgroup.org/public/giant-virus-finder/Giants-in-16Soil-metagenomes}, and the summary of the best hits in Table S3 in the Supplementary material.}
\end{figure}

While the ``Giant Virus Toplist'', defined in the ``methods'' section, contains large phages and a few other viruses that are usually not considered to be Giant viruses, our top results --- measured by E-values and given in Table 1 ---  contains mostly giant viruses when applied to the metagenomes of \cite{Fierer2012}.  For the criterion of assigning a short read to Giant viruses we use a MEGAN5-like approach \cite{Huson2011}: if every taxon in the top-scored 20\% of the Phase 2 alignments are listed in the ``Giant Virus Toplist'', then we accepted the read as a giant virus hit. 

\begin{figure}[t]
	\centering
	\includegraphics[width=5.4in]{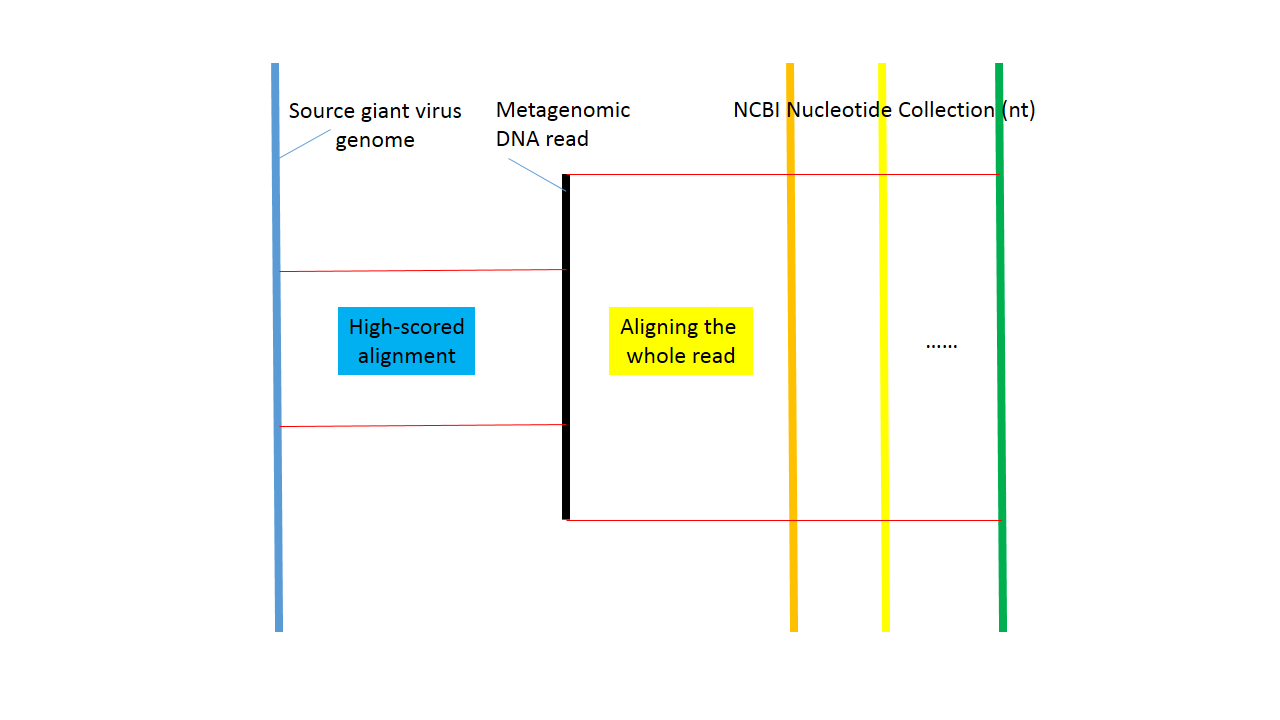}
	\caption{Summary of the Giant Virus Finder workflow. First the giant virus genomes are selected: the selection criterion is a viral genome of size of at least 300,000 bp (if only a partial genome is deposited, its size needs to be at least 300,000 bp). Next, all genomes of giant viruses are aligned to all DNA short reads in the metagenomic dataset. If a high-scored alignment is found, then the {\em whole read} that contains the aligned subsequence (and not only the subsequence of the high-scored alignment) is blasted to the whole NCBI Nucleotide Collection (nt). The short read is accepted as a DNA short read from a giant virus if {\em every sequence} from the top 20\% scored hits, found in the NCBI Nucleotide Collection, corresponds to giant viruses.}
\end{figure}

Samples from Lake Fryxel Valley, Garwood Valley and the Wright Valley, Antarctica, and from Bonanza Creek Forest LTER, Alaska contained the most giant virus taxa. No positive evidence (in the sense described in the ``Methods'' section) was found for the presence of giant virus DNA fragments in the sample originated from the Manu National Park, Peru. 

It is surprising that both hot and cold desert soils contain giant viruses; this finding is in line with our previous result concerning the presence of the giant viruses in the soil samples of the Indian Kutch saline desert \cite{Kerepesi2014b}.

It is worth mentioning that the independent validation of the results presented are easy with the NCBI blastn webserver: one needs to choose a result file which has ``GiantVirusFinder-0.2.fasta" filename ending and then needs to feed it into the NCBI blastn webserver selecting the ``Somewhat similar sequences (blastn)" program option and setting the word size 7 at the ``Algorithm parameters setting" option.

\section{Methods} 

We believe that the method, presented here, is a general workflow: it could also be applied for identifying other sets of taxa, not only giant viruses. The steps of the general workflow:

\begin{itemize}
\item[(\i)] Identify the set $X$ of genomes to be searched for (in our application example $X$ is the set of genomes of the giant viruses); 
\item[(\i\i)] Apply subsequence-search for the sequences in $X$ in the target metagenomic shotgun sequence database $Y$ (in our example $Y$ is one of the 16 metagenomes of \cite{Fierer2012});
\item[(\i\i\i)] Verify the specificity of the hits: the whole fragments in the metagenomic dataset, containing the highest-scored alignments, are aligned to the sequences of a large nucleotide database. Suppose that the top scored hit has score $z$. If all the hits with scores greater than $0.8\times z$ are from the set $X$, ACCEPT, otherwise REJECT the hit (in our example, the hits are aligned to the sequences of the Nucleotide Collection (nt) of the NCBI; and a hit is accepted only if every sequence in the top-scored 20\% belong to set $X$ that is, to the giant virus list).  
\end{itemize}

10\% cut-off  is applied as a default value in the MEGAN phylogenetic analysis tool \cite{Huson2011} for a similar decision. We have found this number is too low for our purpose so we set a more stringent value of 20\%. Users can simply change this threshold. 

The steps of the method are summarized on Figure 2, and on the ``command-by-command level'' on Table S1 and in the README file of the  GiantVirusFinder-1.1.zip archive on \url{http://pitgroup.org/giant-virus-finder/latest}.

\subsection{The Giant Virus Toplist}

In the workflow described above, we need a list $X$ of the genomes and sequences of the organisms we are searching for. 
Defining what is a giant virus and what is not, is a difficult question. We would not like to use potentially questionable and much disputed phylogenetic information in this definition: we simply have constructed the list of viruses with viral genomes or partial genomes (if there is no complete genome deposited) larger than 300 kbp as it is detailed in \url{http://pitgroup.org/giant-virus-toplist/}. Reference genome data are taken from the \url{ftp://ftp.ncbi.nlm.nih.gov/genomes/Viruses/all.fna.tar.gz} file from the NCBI Genome FTP. Note that the length of distinct genome sequences (segments) belonged to a single genome are summarized. Other sequences are added from the NCBI Nucleotide database using the search term: \emph{"Viruses"[Organism] AND 300000:10000000[Sequence Length] NOT "Bacteria"[Organism] NOT "Archaea"[Organism]}.  The list of the viruses found are also given in Table S2 in the supporting material, together with the sequence accession numbers applied in this work.

The inspiration for the Giant Virus Toplist came from \url{http://www.giantvirus.org/top.html}. Our toplist is more up-to-date and contains not only the full-, but also partial genomes.

\subsection{Sequence alignments}

The metagenomic data of the article \cite{Fierer2012} is deposited in the MG-RAST archive:  \url{http://metagenomics.anl.gov/metagenomics.cgi?page=MetagenomeProject\&project=2997}. We downloaded and converted the files into fastq formats. Next, with the stand-alone BLAST distribution \cite{Altschul2005} downloadable {\texttt makeblastdb} program we created 16 BLAST databases for each of the 16 metagenomes.

In Phase 1 (Figure 2) we used the stand alone UNIX blastn program with the default Megablast algorithm changed the word-size from 28 to 16 and  e-value from 10 to 0.01, all the other parameters and the scores and penalties were default for blastn. 

Next, in Phase 2, the hits with better E-value than 0.01 were collected from each alignment, and were aligned using \texttt blastn with word-size of 7 against the whole Nucleotide Collection (nt) of the NCBI. Suppose that the top scored hit has score $z$. If all the hits with scores greater than $0.8\times z$ are from the The Giant Virus Toplist, we accepted the hit, otherwise rejected it. 

The summary of the results of the two-phase search process with the highest scored giant viruses are given in the Supplementary Table S3, their numbers in Figure 1 and Figure S1. All the files created by the workflow are given at \url{http://pitgroup.org/public/giant-virus-finder/Giants-in-16Soil-metagenomes/}.

\subsection{The advantage of the two-phase method}

Using a straightforward one phase method (simply blastn all reads against the nt database with the word-size=7 option) would require about 1080 years (about 0,084 h/read) in a machine using a single CPU core.  Selecting 9,829 candidate reads from the whole 112,674,624 reads of the 16 metagenomes in Phase 1 reduced the running time to about 34 days in a single-core machine.

\bigskip
{\noindent \bf Data availability:} 
The metagenomes of the article \cite{Fierer2012} can be downloaded from: \url{http://metagenomics.anl.gov/metagenomics.cgi?page=MetagenomeProject\&project=2997}. The Giant Virus Finder is downloadable from \url{http://pitgroup.org/public/giant-virus-finder/latest}. The detailed alignment results in both phases of the search are found in \url{http://pitgroup.org/public/giant-virus-finder/Giants-in-16Soil-metagenomes}.

\section{Conclusions} 
We have shown the very probable presence of giant viruses in diverse environmental soil samples by a two-phase search strategy in metagenomic samples and the NCBI Nucleotide Collection (nt). Consequently, such non-aqueous environments as Antarctic dry valleys, the Mojave desert, the prairie and several forest-soils most probably also contain these recently discovered viruses. It is a surprising result that we have found an abundance of giant viruses in samples from Antarctic dry valleys.

\section{Acknowledgements:} The authors declare no conflicts of interest.

\section{References}


\end{document}